\documentclass{ifacconf}

\usepackage{graphicx}      
\usepackage{natbib}       
\usepackage[dvipsnames]{xcolor}
\usepackage{enumerate}
\usepackage{enumitem}
\usepackage{amsmath}
\usepackage{amssymb}
\usepackage{amsfonts}
\usepackage{dsfont}
\usepackage{appendix}
\setcitestyle{square}

\newtheorem{theorem}{Theorem}[section]

\newtheorem{lemma}{Lemma}[section]
\newtheorem{remark}{Remark}

\newtheorem{assumption}{Assumption}

\newtheorem{definition}{Definition}

\def\bs{\boldsymbol}
\def\mbb{\mathbb}
\def\mcal{\mathcal}
\def\mca{\mathcal{A}}
\def\mcb{\mathcal{B}}

\def\F{\mathbb{F}}
\def\E{\mathbb{E}}
\def\P{\mathbb{P}}

\begin{document}
\begin{frontmatter}

\title{Strategically revealing capabilities in General Lotto games\thanksref{footnoteinfo}} 

\thanks[footnoteinfo]{This work was supported in part by NSF Award \#ECCS-2013779.}

\author[First]{Keith Paarporn} 
\author[First]{Philip N. Brown} 

\address[First]{University of Colorado, Colorado Springs, 
   Colorado Springs, CO 80918 USA (e-mail: kpaarpor@uccs.edu, pbrown2@uccs.edu).}

\begin{abstract}                

Can revealing one's competitive capabilities to an opponent offer strategic benefits? In this paper, we address this question in the context of General Lotto games, a class of two-player competitive resource allocation models. We consider an asymmetric information setting where the opponent is uncertain about the resource budget of the other player, and holds a prior belief on its value. We assume the other player, called the signaler, is able to send a noisy signal about its budget to the opponent. With its updated belief, the opponent then must decide to invest in costly resources that it will deploy against the signaler's resource budget in a General Lotto game. We derive the subgame perfect equilibrium to this extensive-form game. In particular, we identify necessary and sufficient conditions for which a signaling policy improves the signaler's resulting performance in comparison to the scenario where it does not send any signal. Moreover, we provide the optimal signaling policy when these conditions are met. Notably we find that for some scenarios, the signaler can effectively double its performance.

\end{abstract}


\end{frontmatter}

\section{Introduction}

The advancement of communication technologies and platforms have fundamentally shifted how information is sent, perceived, and ultimately utilized to make decisions. From a system-level perspective, revealing, concealing, or manipulating information can be a viable and impactful method for the control of multi-agent systems, particularly when those systems include strategic decision-makers.
For example, travel information systems may broadcast recommended routes to drivers in a transportation network with the objective of lowering overall congestion~\citep{Zhu2020}.
Advertisers may influence user behavior (e.g. purchasing decisions, engaging with certain content) on online platforms by making personalized recommendations~\citep{Ke2022,candogan2020optimal}. 
In these systems, a central authority implements a signaling policy that carefully determines what information to broadcast to a collection of uninformed users.

Due to its wide-ranging potential applications, the problem of strategically revealing or concealing information has received a great deal of research attention in recent years, often under the name of \textit{Bayesian persuasion}~\citep{kamenica2011bayesian} or \emph{information design}~\citep{Bergemann2019}.
Under these frameworks, an informed agent crafts the information that is revealed to an uninformed agent (or agents); by selecting the revelation policy carefully, the informed agent can influence the posterior beliefs of the uninformed agent and can often thus control its resulting decisions. These techniques have primarily been applied to study the influence of multi-agent systems such as transportation networks, epidemic management, and social networks \citep{Massicot2019,Zhu2020,liu2022eproach,wu2019information,candogan2020optimal}. 

In this paper, we focus on an information designer (called the \emph{signaler}) that has an opportunity to reveal information about its own competitive capabilities to an \emph{adversary}, with which it is in direct competition. We study such a scenario in the context of General Lotto games, a popular game-theoretic model of competitive resource allocation. We base our analysis on a General Lotto game where the resource budget of the signaler is randomly drawn from a publicly-known Bernoulli distribution (studied recently in \cite{Paarporn_2021_budget}). The true budget is the private information of the signaler. The signaler adopts a \emph{signaling policy}; once the signaler's resource budget is realized, the signaling policy delivers a noisy signal of its budget to the adversary.
The adversary then updates its beliefs about the signaler's budget, chooses an amount of resources to invest in (which is publicly disclosed), and subsequently engages in a General Lotto game with the signaler using its invested resources. We are primarily concerned with deriving the signaler's optimal policy, and identifying conditions on environmental parameters for which it offers performance improvements compared to not signaling at all. We note that our study is a departure from the traditional Bayesian Persuasion setup, for which the signaler does not participate in strategic interactions after signaling to the receiver.

Our main results are given as follows.
\begin{itemize}
    \item We fully characterize the signaler's optimal signaling policy in all instances of the game.
    \item We fully characterize necessary and sufficient conditions for which the optimal signaling policy provides performance improvements for the signaler.
\end{itemize}
We find that the optimal signaling policy derived maximizes the probability that the signal sent deters the adversary from competing at all in the General Lotto game. Interestingly, we find there are parameters under which the signaler can effectively double its performance by using an optimal policy.


\noindent\textbf{Related works:} Much recent work has been devoted to strategic signaling, particularly in the area of cyber-physical-human systems such as transportation networks \citep{Massicot2019,Wu2021,Gould2022a,Ferguson2022}. In these works, a system planner who is informed about the state of the world (e.g., the presence of traffic accidents on a highway) decides what information to transmit to a user or group of uninformed self-interested system users, with the typical goal of improving system performance.
Thus, this line of research is broadly focused on using strategic information provision in a benevolent way to act as a coordination mechanism among a large population of disorganized decision-makers.
In contrast, our paper studies the information design problem in a competitive setting; i.e., rather than attempting to coordinate behavior among a group of users, our signaler is attempting to manipulate the uncertainty of an adversary.

Another line of research studies information design in competitive settings, but focuses mainly on competition in markets: e.g., two firms competing over market share may reveal/conceal information about their competitors' product quality~\citep{kamenica2011bayesian,Ivanov2013,Board2018,Li2018}.
These works generally focus on the effects of competitive information provision on metrics such as consumer surplus.
In contrast, our model considers providing information directly to an adversary, rather than to market participants.

Perhaps most closely aligned with our work is the recent studies on strategic information provision in \textit{contests}. Several recent works have shown that the strategic revelation of information to opponents can serve as a viable competitive strategy, ranging from pre-commitments to full revelation of information \citep{paarporn2021strategically,Epstein2013}. \cite{Zhang2016} consider a scenario in which a \emph{contest organizer} can force contestants to reveal private information in an attempt to maximize the effort expended by contestants; in that work, if contestants have binary valuations, the optimal policy is either no or full disclosure.
Other papers study similar problems;~\cite{Fu2011} study disclosure policy when contest entry is stochastic and~\cite{Denter2011} examine the strategic effects of time-delayed information revelation.
Similar to our work (but using a different contest model),~\cite{Epstein2013} consider a contest in which the informed player can choose to either conceal or fully reveal its abilities to the uninformed player.

In this paper, we consider the role that information signaling has in adversarial interactions. The driving question is: can signaling one's capabilities to an uncertain opponent offer strategic benefits?

\section{Preliminaries on General Lotto games}


To build up to our signaling General Lotto game, we provide some background on two-player simultaneous-move General Lotto games. First, we present the classic complete information formulation in Section \ref{sec:CI_lotto}. Then we present an incomplete information setting with asymmetric budget uncertainty from the recent literature \cite{} in Section \ref{sec:BL_lotto}.

\subsection{Complete information Lotto games}\label{sec:CI_lotto}

A (complete information) General Lotto game consists of two players, $\mca$ and $\mcb$. Each player is tasked with allocating their endowed resource budgets $A, B > 0$ across a set of $n$ battlefields. Each battlefield has an associated value $v_j > 0$, $j\in[n]$.  An allocation for $\mca$ is any vector $\bs{x}_{\mca} \in \mbb{R}_{\geq 0}^n$, and similarly for $\mcb$. An admissible strategy for $\mca$ is a randomization $F_\mca$ over allocations such that the expended resources do not exceed the budget $A$ in expectation. Specifically, $F_\mca$ is an $n$-variate (cumulative) distribution that belongs to the family
\begin{equation}\label{eq:LC}
    \F(A) \triangleq \left\{ F : \E_{\bs{x}_{\mca}\sim F}\left[\sum_{j=1}^n x_{\mca,j}\right] \leq A \right\}.
\end{equation}
and similarly, $F_\mcb \in \F(B)$. Given a strategy profile $(F_\mca,F_\mcb)$, the utility of player $\mca$ is 
\begin{equation}\label{eq:Lotto_payoff}
    u_\mca(F_\mca,F_\mcb) \triangleq \E_{\substack{\bs{x}_{\mca}\sim F_\mca \\ \bs{x}_{\mcb}\sim F_\mcb}}\left[\sum_{j=1}^n v_j \cdot  \mathds{1}_{\{x_{\mca,j} \geq x_{\mcb,j} \}} \right]
\end{equation}
where $\mathds{1}_{\{\cdot\}}$ is 1 if the statement in the bracket is true, and 0 otherwise\footnote{An arbitrary tie-breaking rule may be selected, without changing our results. This is generally true in General Lotto games \citep{Kovenock_2021}. For simplicity, we will assume ties are awarded to player $\mca$.}. It follows that the utility of player $\mcb$ is
\begin{equation}
    u_\mcb(F_\mca,F_\mcb) \triangleq \phi - u_\mca(F_\mca,F_\mcb)
\end{equation}
where $\phi \triangleq \sum_{j=1}^n v_j$ is the total sum of battlefield values. An instance of the complete information General Lotto game is denoted by $\text{GL}(A,B)$. An equilibrium is a strategy profile $(F_\mca^*,F_\mcb^*)$ such that
\begin{equation}
    \begin{aligned}
        u_\mca(F_\mca^*,F_\mcb^*) &\geq u_\mca(F_\mca,F_\mcb^*), \quad \forall F_\mca \in \F(A) \\
        u_\mcb(F_\mca^*,F_\mcb^*) &\geq u_\mcb(F_\mca^*,F_\mcb), \quad \forall F_\mcb \in \F(B).
    \end{aligned}
\end{equation}

The unique equilibrium payoffs in General Lotto games is well-established in the literature.

\begin{theorem}[\cite{Kovenock_2021}]
    Consider any General Lotto game $\text{GL}(A,B)$. The payoff to player $\mca$ in any equilibrium is given by
    \begin{equation}
        \pi_\mca^\text{ci}(A,B) \triangleq \phi\cdot
        \begin{cases}
            \frac{A}{2B}, &\text{if } A \leq B \\
            1 - \frac{B}{2A}, &\text{if } A > B
        \end{cases},
    \end{equation}
    and the payoff to player $\mcb$ is $\pi_B^\text{ci}(A,B) = \phi - \pi_A^\text{ci}(A,B)$.
\end{theorem}
Note that the payoffs depend on the total value $\phi$, and not on individual values of the valuation vector $\bs{v}$. Hence, we omit specifying $\bs{v}$ in the notation $\text{GL}(A,B)$.

\subsection{Lotto games with asymmetric budget uncertainty}\label{sec:BL_lotto}

\begin{figure*}[t]
    \centering
    \includegraphics[scale=0.22]{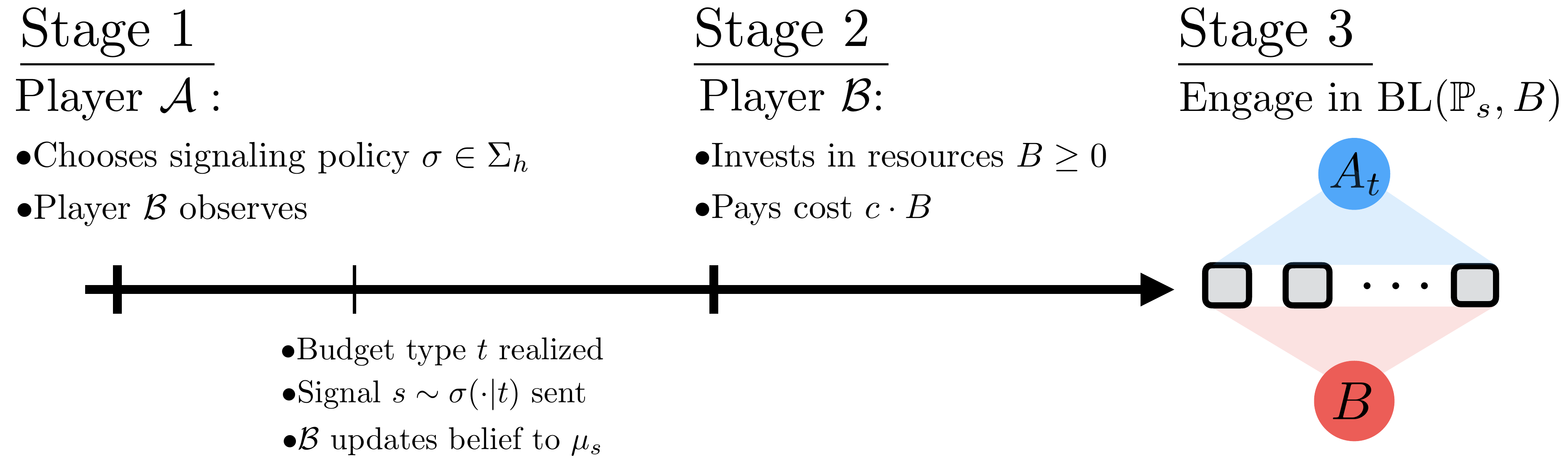}
    \caption{\small The signaling General Lotto game. In Stage 1, the signaler (player $\mca$) selects a signaling policy $\sigma$, which becomes common knowledge. Before Stage 2, the signaler's budget type is realized according to the prior distribution $p$, and a signal $s$ is sent to player $\mcb$ according to the distribution specified by $\sigma$. The receiver (player $\mcb$) updates its belief. In Stage 2, player $\mcb$ decides how many resources $B \geq 0$ to invest in, paying the cost $c\cdot B$. In Stage 3, both players simultaenously engage against each other in the Bernoulli Lotto game. The sequence of choices and events is standard in the information design literature \citep{kamenica2019bayesian}. }
    \label{fig:my_label}
\end{figure*}

We present an asymmetric information General Lotto game from the recent literature \citep{Paarporn_2021_budget}, which will serve as the basis of our signaling General Lotto game. 

Player $\mca$ has two possible budget types, $t \in T = \{h,\ell\}$. The type is drawn according to a Bernoulli distribution with parameter $p$.  Specifically, with probability $p\in[0,1]$, $\mca$'s type is $h$ and it is endowed with a high budget $A_h$. With probability $1-p$, $\mca$'s type is $\ell$ and it is endowed with a low budget $A_\ell$, with $A_h \geq A_\ell \geq 0$. The realized type is known only to player $\mca$, but the distribution and values $A_h,A_\ell$ are common knowledge. Player $\mcb$ thus holds the prior belief $p$ on the high type $A_h$. The budget endowment $B \geq 0$ of player $\mcb$ is common knowledge.

An admissible strategy for player $\mca$ is a pair $\vec{F}_\mca = \{F_\mca^h,F_\mca^\ell \} \in \F(A_h)\times\F(A_\ell)$, where $F_\mca^h$ or $F_\mca^\ell$ is implemented depending on which budget type is realized. An admissible strategy for player $\mcb$ is a single strategy $F_\mcb \in \F(B)$ that is implemented regardless of which type is realized. If $\mca$'s private type is $t \in \{h,\ell\}$, the ex-interim expected utilities given the strategy profile $(\vec{F}_\mca,F_\mcb)$ are defined as
\begin{equation}
    \begin{aligned}
        &U_\mca(\vec{F}_\mca,F_\mcb;t) \triangleq u_\mca(F_\mca^t,F_\mcb) \\
        &U_\mcb(F_\mcb,\vec{F}_\mca) \triangleq p\cdot u_\mcb(F_\mcb,F_\mca^h) + (1-p)\cdot u_\mcb(F_\mcb,F_\mca^\ell)
    \end{aligned}
\end{equation}
where $u_\mca$ and $u_\mcb$ are defined from \eqref{eq:Lotto_payoff}. A Bayes-Nash equilibrium is a strategy profile $(\vec{F}_\mca^*,F_\mcb^*)$ such that for every type $t \in \{h,\ell\}$,
\begin{equation}\label{eq:BNE}
    \begin{aligned}
        U_\mca(\vec{F}_\mca^*,F_\mcb^*;t) &\geq U_\mca(\vec{F}_\mca,F_\mcb^*;t), \hspace{1mm} &\forall \vec{F}_\mca \in \F(A_h)\times\F(A_\ell) \\
        U_\mcb(F_\mcb^*,\vec{F}_\mca^*) &\geq U_\mcb(F_\mcb,\vec{F}_\mca^*),  &\forall F_\mcb \in \F(B)
    \end{aligned}
\end{equation}
We refer to this setup as a Bernoulli Lotto game, where a particular instance is characterized by the parameters $\P \triangleq (A_h,A_\ell,p)$ and $B$. We will denote a Bernoulli Lotto game as $\text{BL}(\P,B)$.

Unique equilibrium payoffs to every Bernoulli Lotto game were completely characterized in recent work \citep{Paarporn_2021_budget}. For an equilibrium $(\vec{F}_\mca^*,F_\mcb^*)$, we will write player $\mca$'s equilibrium ex-interim payoff in $\text{BL}(\P,B)$ given the budget type $t \in \{h,\ell\}$ as 
\begin{equation}
    \pi_\mca(\P,B;t) \triangleq U_\mca(\vec{F}_\mca^*,F_\mcb^*;t)
\end{equation}
Player $\mcb$ has a single (public) budget type and must reason (based on its belief) about the budget type of player $\mca$. Its equilibrium payoff is written as
\begin{equation}
    \pi_\mcb(\P,B) \triangleq \phi - (p\pi_\mca(\P,B;h) + (1-p)\pi_\mca(\P,B;\ell) ).
\end{equation}


\section{The signaling General Lotto game}

Consider a Bernoulli Lotto game. Before engaging in competition, player $\mca$ has an opportunity to re-shape the belief of player $\mcb$ by sending a (noisy) signal $s \in \{h,\ell\}$ directly to player $\mcb$. Specifically, player $\mca$ adopts a \emph{signaling policy} $\sigma = \{\sigma(s | t)\}_{s\in S, t\in T}$, where $S = \{h,\ell\}$ is the set of possible signals that can be sent to player $\mcb$. Here, $\sigma(s|t)$ is the probability that the signal $s \in S$ is sent to player $\mcb$, given that budget type $t \in T$ is realized. 

In this paper, we will focus on the sub-class of signaling policies that truthfully signal `high' when the budget type is actually high. Such an assumption is warranted -- intuitively, a competitor would not want to signal that it is weaker than it actually is. Future work will analyze the scenario where this assumption is eliminated\footnote{Computational results (not shown) indeed suggest that even when Assumption 1 is eliminated, the same SPE signaling policy for player $\mca$ detailed in the main result, Theorem \ref{thm:signal_regions}, still holds.}.

\begin{assumption}\label{assump:high}
    Player $\mca$'s set of admissible signaling policies is restricted to the sub-class of signaling policies $\sigma \in \Sigma_h$, where $\Sigma_h$ is the set of policies that satisfy $\sigma(h|h) = 1$.
\end{assumption}

Any signaling policy in this class $\sigma \in \Sigma_h$ satisfies
\begin{equation}
    \begin{aligned}
        &\sigma(h|h) = 1, \quad \sigma(\ell | h) = 0 \\
        &\sigma(h|\ell) + \sigma(\ell|\ell) = 1
    \end{aligned}
\end{equation}

Note that any signaling policy $\sigma \in \Sigma_h$ is completely characterized by a single number: $q = \sigma(h|\ell)$. 


The interaction unfolds in the following three-stage extensive form game. 

\noindent\textbf{Stage 1: } Player $\mca$ selects a feasible signaling policy $\sigma \in \Sigma_h$. The budget type $t \in T$ is drawn, and a signal $s \sim \sigma(\cdot|t)$ is sent to player $\mcb$. Player $\mcb$ performs a Bayesian update on its prior belief. Specifically, if $s = h$, $\sigma$ induces the posterior belief $\mu_h \triangleq\frac{p}{p + q (1-p)}$ on the high budget type. If $s = \ell$, $\sigma$ induces the posterior belief $\mu_\ell \triangleq 0$ on the high budget type.


\noindent\textbf{Stage 2: } Player $\mcb$ selects an amount of resources $B \geq 0$ to invest in. It pays the cost $c\cdot B$, where $c > 0$ is its \emph{per-unit cost}.

\noindent\textbf{Stage 3: } Player $\mca$ (with budget $A_t$) and player $\mcb$ (with budget $B$) engage in the ex-interim stage of the Bernoulli Lotto game $\text{BL}(\P_s,B)$, where $\P_s \triangleq (A_h,A_\ell,\mu_s)$ ($s \in \{h,\ell\}$) is $\mcb$'s posterior distribution on budget types. The final payoff to $\mca$ is its ex-interim equilibrium payoff in $\text{BL}(\P_s,B)$, denoted $\pi_\mca(\P_s,B;t)$. The final payoff to $\mcb$ is its (Bayes-Nash) equilibrium payoff under the belief $\P_s$ (denoted $\pi_\mcb(\P_s,B)$) minus the investment cost $c\cdot B$ from Stage 2.

The prior distribution on budget types $\P = (A_h,A_\ell,p)$ and player $\mcb$'s per-unit cost $c$ are defining parameters in the extensive-form signaling game. We denote an instance of the game as $\text{SG}(\P,c)$. We consider the following standard solution concept for extensive-form games.

\begin{definition}
    A pair $(\sigma^*,B^*)$, where $\sigma^* \in \Sigma_h$ and $B^* : [0,1] \rightarrow \mbb{R}_{\geq 0}$, is a \emph{subgame perfect equilibrium (SPE)} of $\text{SG}(\P,c)$ if:
    \begin{enumerate}
        \item For any signaling policy $\sigma \in \Sigma_h$ and $s \sim \sigma(\cdot|t)$ with $t \in T$,
        \begin{equation}\label{eq:optimal_B_problem}
            B^*(\mu_s) \in \arg\max_{B\geq 0} \{ \pi_\mcb(\P_s,B) - c\cdot B\}
        \end{equation}
        \item The signaling policy $\sigma^*$ solves
        \begin{equation}\label{eq:optimal_sigma_problem}
            \max_{\sigma \in \Sigma_h} \Pi_\mca(\sigma) \triangleq \E_{t}\left[\E_{s\sim \sigma(\cdot|t)} \left[\pi_\mca(\P_s,B^*(\mu_s);t) \right] \right].
        \end{equation}
    \end{enumerate}
\end{definition}
Note that player $\mcb$'s investment decision in Stage 2 is contingent on a realization of the signal $s$ \eqref{eq:optimal_B_problem}. Player $\mca$'s choice of signaling policy $\sigma$ in Stage 1 is taken before its budget type is realized -- it thus considers final payoffs in expectation with respect to $t$ and $s$ \eqref{eq:optimal_sigma_problem}. These are standard formulations in the Bayesian persuasion literature \citep{kamenica2019bayesian}.

\begin{definition}
    The \emph{trivial signaling policy}, denoted $\varnothing$, is the signaling policy for which $q=1$. In particular, this policy always sends $s=h$ regardless of the budget type, leaving player $\mcb$'s posterior belief unchanged from the prior, i.e. $\mu_h = \mu_\ell = p$. We denote the payoff player $\mca$ obtains by implementing a trivial policy as $\Pi_\mca^\text{ns} \triangleq \Pi_\mca(\varnothing)$.
\end{definition}


The trivial signaling policy is thus equivalent to not signaling at all. The primary goal of this paper is to identify conditions on $(\P,c)$ for which player $\mca$'s payoff in an SPE of $\text{SG}(\P,c)$ exceeds  $\Pi_\mca(\varnothing)$.

\section{Main results}


The following result identifies necessary and sufficient conditions for which the signaling policy from the SPE outperforms the trivial policy.

\begin{figure}
    \centering
    \includegraphics[scale=0.22]{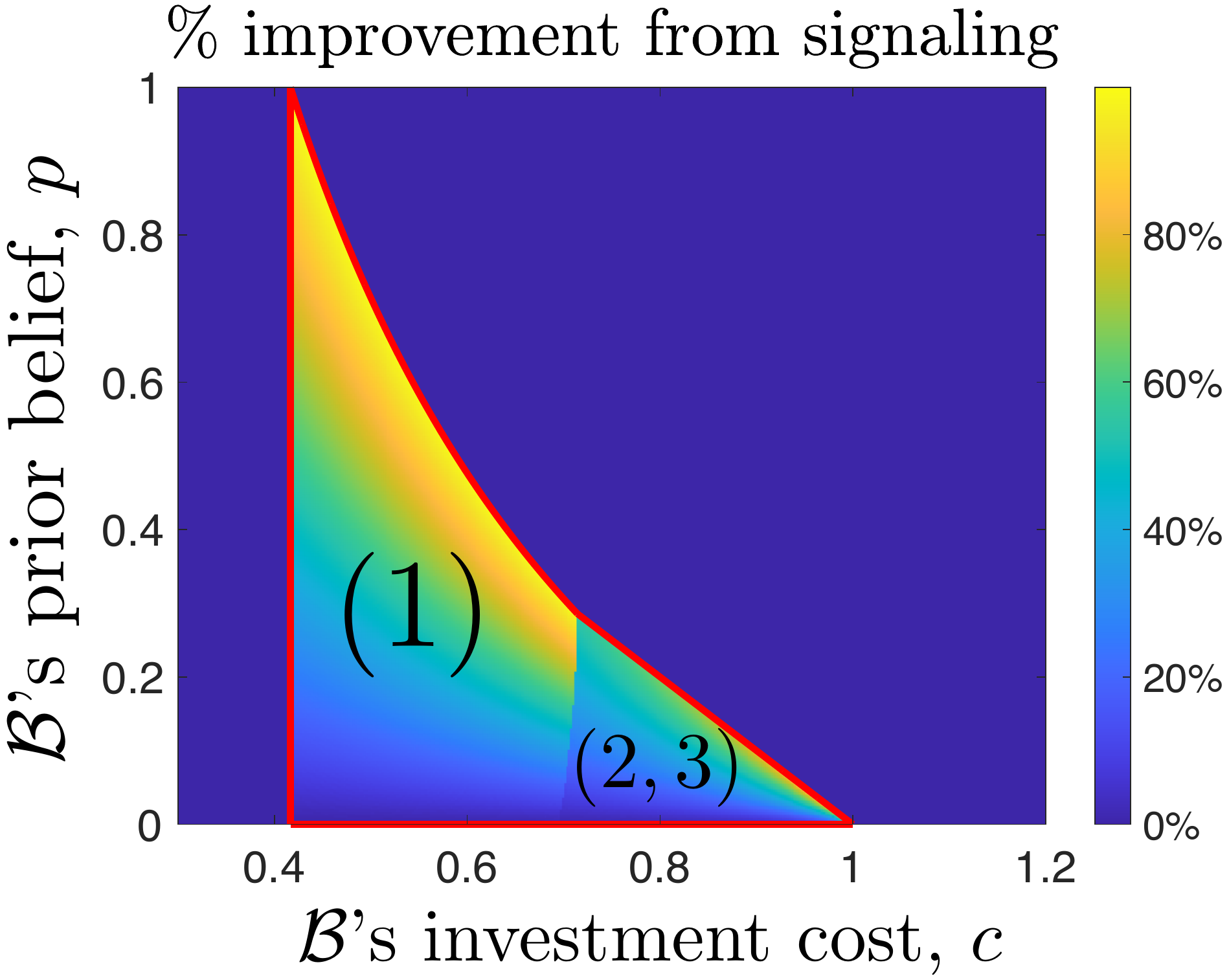}
    \caption{\small Illustration of the range of parameters where signaling is beneficial for player $\mca$. The labels indicate the regions specified in Theorem \ref{thm:signal_regions}. The intensity of the plot indicates the value $100\times(\frac{\Pi_\mca^*}{\Pi_\mca^\text{ns}} - 1)$, which is the percent improvement attainable from the SPE signaling policy. In this plot, we have set $A_h = 1.2$, $A_\ell = 0.5$, and $\phi = 1$.}
    \label{fig:signal_regions}
\end{figure}

\begin{theorem}\label{thm:signal_regions}
    The SPE signaling policy $\sigma^*\in\Sigma_h$ of $\text{SG}(\P,c)$ outperforms the trivial policy, i.e. $\Pi_\mca(\sigma^*) > \Pi_\mca^\text{ns}$, if and only if $(c,p)$ satisfies either of the following:
    \begin{enumerate}
        \item $\frac{\phi}{2A_h} \leq c < \min\left\{\frac{\phi}{2\bar{A}(p)}, \lambda(p) \right\}$ and $p \in [0,1]$, where 
        \begin{equation}\label{eq:lambda}
            \lambda(p) \triangleq \frac{\phi}{2A_h^2}\left(\sqrt{(1-p)A_\ell} + \sqrt{pA_h + (1-p)A_\ell}\right)^2.
        \end{equation}
        The SPE signaling policy is  $\sigma^*(h|\ell) = \frac{p}{1-p}\frac{2cA_h-\phi}{\phi-2cA_\ell}$.
        \item $\max\{\lambda(p),\frac{\phi}{2A_h}\} \leq c < \frac{\phi}{2(A_h-A_\ell)}$, $p < \frac{A_h - 2A_\ell}{A_h - A_\ell}$, and 
        \begin{equation}\label{eq:case2_condition}
            f_p > \frac{\phi-2c A_\ell}{2cA_h-\phi} \cdot \frac{\sqrt{c\phi A_\ell/2}}{\phi - \sqrt{c\phi A_\ell/2}}
        \end{equation}
        where $f_p \triangleq \frac{p}{\sqrt{1-p} - (1-p)}$. The SPE signaling policy is $\sigma^*(h|\ell) = \frac{p}{1-p}\frac{2cA_h-\phi}{\phi-2cA_\ell}$.
        \item $\max\left\{ \frac{\phi}{2(A_h-A_\ell)}, \frac{\phi}{2A_\ell}\left(\frac{f_p - \sqrt{f_p^2 - f_p + 1}}{f_p-1} \right)^2 \right\} \leq c < \frac{(1-p)\phi}{2A_\ell}$ and $p < \frac{A_h - 2A_\ell}{A_h - A_\ell}$. The SPE signaling policy is $\sigma^*(h|\ell) = \frac{p}{1-p}\frac{2cA_\ell}{\phi-2cA_\ell}$.
    \end{enumerate}
\end{theorem}
Figure \ref{fig:signal_regions} depicts an example of the regions given by the above conditions, as well as the percent improvement in $\mca$'s performance that it obtains by implementing the SPE signaling policy.

\noindent\emph{Discussion of main results}: 
There are several interesting observations of Theorem \ref{thm:signal_regions}. For parameters near the top border of the region  in Figure \ref{fig:signal_regions} (i.e. setting $c = \frac{\phi}{2A_h}$ and $p \rightarrow 1$), player $\mca$'s performance approaches a two-fold improvement compared to the performance of the trivial policy. Additionally, the sharp discontinuities in the plot indicate that player $\mca$'s performance, and in turn, player $\mcb$'s investment decision, is highly sensitive to changes in the underlying parameters. 

Outside of the indicated regions, the SPE signaling policy is the trivial policy. That is, no signaling policy can strictly improve upon $\Pi_\mca^\text{ns}$. For parameters to the right of the indicated regions, player $\mcb$'s cost to invest in resources is sufficiently expensive such that its SPE investment is zero when $\mca$ uses the trivial signaling policy. In particular, player $\mca$ is able to win the entire contest without signaling at all. Now, consider parameters to the left of the indicated regions. Player $\mcb$'s cost to invest in resources is cheap enough such that its SPE investment is high, and no signaling policy is able to induce $\mcb$ to invest in a lower amount of resources.

Under Assumption \ref{assump:high}, a signaling policy $\sigma\in\Sigma_h$ is determined by a single number $\sigma(h|\ell) = q \in [0,1]$, i.e. the probability the signal $s=h$ is sent when the budget type is low. The SPE policy is given by $q^*$ (values specified in the statement). Interestingly, when the parameters satisfy any of the conditions in Theorem \ref{thm:signal_regions}, the policies in the range $q \in [0,q^*]$ induce player $\mcb$ to invest zero resources in Stage 2, given that it received the signal $s = h$. In a sense, $q^*$ is intuitively the highest fraction of time $\mca$ can ``lie" about its low budget type, such that player $\mcb$ will refrain from competing in the General Lotto game in Stage 3. Moreover, we find that for condition (1) in Theorem \ref{thm:signal_regions}, the signaling policy $q=0$ still provides an improvement over a trivial policy. In other words, full revelation of one's budget type performs better than not revealing any additional information for certain parameters.


\section{Analysis}


In this section, we provide the derivation of the SPE of $\text{SG}(\P,c)$, and the proof of Theorem \ref{thm:signal_regions}. The analysis hinges on using the equilibrium characterizations from previous work \citep{Paarporn_2021_budget}. We first provide the benchmark payoff that player $\mca$ obtains with a trivial policy. 

\begin{lemma}[Lemma C.2 in \citep{Paarporn_2021_budget}] \ 

    Consider the extensive form game $\text{SG}(\P,c)$. If $p \geq \frac{A_h - 2A_\ell}{A_h - A_\ell}$, then 
    \begin{equation}\label{eq:Pi_ns_1}
        \Pi_\mca^\text{ns} = 
        \begin{cases}
            \sqrt{\frac{c\phi\bar{A}(p)}{2}}, &\text{if } c < \frac{\phi}{2\bar{A}(p)} \\
            \phi, &\text{if } c \geq \frac{\phi}{2\bar{A}(p)}
        \end{cases}.
    \end{equation}
    If $p < \frac{A_h - 2A_\ell}{A_h - A_\ell}$, then
    \begin{equation}\label{eq:Pi_ns_2}
        \Pi_\mca^\text{ns} = 
        \begin{cases}
    	    \sqrt{\frac{c\phi\bar{A}(p)}{2}}, &\text{if } c \in [0, \lambda(p)) \\
    		p\phi  + \sqrt{\frac{c\phi (1-p)A_\ell}{2}}, &\text{if } c \in [\lambda(p),\frac{(1-p)\phi}{2A_\ell}) \\
    		\phi, &\text{if } c \geq \frac{(1-p)\phi}{2A_\ell}
	    \end{cases}
    \end{equation}
    where $\bar{A}(p) \triangleq p A_h + (1-p)A_\ell$ is the expected budget of $\mca$ under belief $p$, and $\lambda(p)$ was defined in \eqref{eq:lambda}.
\end{lemma}


\subsection{SPE investment level}

The SPE of $\text{SG}(\P,c)$ can be derived by backwards induction. Hence, we first derive the optimal investment $B^*(\mu)$ \eqref{eq:optimal_B_problem} in Stage 2, given any belief $\mu \in [0,1]$ on the high budget. The characterization of $B^*(\mu)$ is derived from prior work:
\begin{lemma}[Lemma C.1 in \citep{Paarporn_2021_budget}]\label{lem:Bstar} \ 

    Consider any posterior belief $\mu \in [0,1]$. If $\mu \geq \frac{A_h - 2A_\ell}{A_h - A_\ell}$, then the SPE investment for player $\mcb$ is
    \begin{equation}\label{eq:Bstar_1}
        B^*(\mu) = 
        \begin{cases}
            \sqrt{\frac{ \bar{A}(\mu)\phi}{2 c}}, &\text{if } c < \frac{\phi}{2\bar{A}(\mu)} \\
            0, &\text{if } c \geq \frac{\phi}{2\bar{A}(\mu)}
        \end{cases}.
    \end{equation}
    If $\mu < \frac{A_h - 2A_\ell}{A_h - A_\ell}$, then the SPE investment for player $\mcb$ is
    \begin{equation}\label{eq:Bstar_2}
		B^*(\mu) = 
		\begin{cases}
			\sqrt{\frac{\bar{A}(\mu) \phi}{2c}}, &\text{if } c \in [0, \lambda(\mu)) \\
			\sqrt{\frac{(1-\mu)A_\ell\phi}{2c}}, &\text{if } c \in [\lambda(\mu),\frac{(1-p)\phi}{2A_\ell}) \\
			0, &\text{if } c \geq \frac{(1-\mu)\phi}{2A_\ell}
		\end{cases}
	\end{equation}
	where $\lambda(\mu)$ is defined in \eqref{eq:lambda}.
\end{lemma}

Here, we observe that the SPE investment $B^*(\mu)$ is increasing in the belief $\mu$ for low investment costs (first entry of \eqref{eq:Bstar_1}, \eqref{eq:Bstar_2}), is decreasing for intermediate investment costs (second entry of \eqref{eq:Bstar_2}), and is zero for high investment costs (second entry of \eqref{eq:Bstar_1}, third entry of \eqref{eq:Bstar_2}).

\subsection{SPE signaling policy}

Under Assumption \ref{assump:high}, any signaling policy $\sigma \in \Sigma_h$ is uniquely determined by a single variable $q \triangleq \sigma(h|\ell)$. For any $q \in [0,1]$, we can write the objective in \eqref{eq:optimal_sigma_problem} as
\begin{equation}\label{eq:final_payoff_Sigmah}
    \begin{aligned}
        \Pi_\mca(\sigma) &= p \cdot\pi_\mca(\P_h,B^*(\mu_h);h) \\
        &+ (1-p)\cdot q \cdot \pi_\mca(\P_h,B^*(\mu_h);\ell) \\
        &+  (1-p)\cdot(1-q)\cdot\pi_\mca(\P_\ell,B^*(\mu_\ell);\ell)
    \end{aligned}
\end{equation}
where under Assumption \ref{assump:high}, $\mu_h = \frac{p}{p + q(1-p)}$ and $\mu_\ell = 0$. Using Lemma \ref{lem:Bstar} and the characterizations $\pi_\mca$ from \citep{Paarporn_2021_budget}, we obtain the following expressions for player $\mca$'s ex-interim equilibrium payoffs $\pi_\mca(\P_s,B^*(\mu_s);t)$.

\begin{lemma}[\citep{Paarporn_2021_budget}]\label{lem:A_ex_interim}
    Consider any posterior belief $\mu\in[0,1]$, and denote $\P_\mu = (A_h,A_\ell,\mu)$. If $\mu \geq \frac{A_h - 2A_\ell}{A_h - A_\ell}$, then for $t \in \{h,\ell\}$:
    \begin{equation}\label{eq:Pi_At_1}
        \pi_\mca(\P_\mu,B^*(\mu);t) = 
        \begin{cases}
            \sqrt{\frac{c\phi A_t^2}{2 \bar{A}(\mu)}}, &\text{if } c < \frac{\phi}{2\bar{A}(\mu)} \\
            \phi, &\text{if } c \geq \frac{\phi}{2\bar{A}(\mu)}
        \end{cases}.
    \end{equation}
    If $\mu < \frac{A_h - 2A_\ell}{A_h - A_\ell}$, then for $t \in \{h,\ell\}$:

    \begin{equation}\label{eq:Pi_At_2}
        \begin{aligned}
            &\pi_\mca(\P_\mu,B^*(\mu);t) = \\
            &\begin{cases}
        	    \sqrt{\frac{c\phi A_t^2}{2\bar{A}(\mu)}}, &\text{if } c \in [0, \lambda(\mu)) \\
        		\mathds{1}_{t=h}\cdot \phi  + \mathds{1}_{t=\ell}\cdot\sqrt{\frac{c\phi A_\ell}{2(1-\mu)}}, &\text{if } c \in [\lambda(\mu),\frac{(1-\mu)\phi}{2A_\ell}) \\
        		\phi, &\text{if } c \geq \frac{(1-\mu)\phi}{2A_\ell}
    	    \end{cases}
        \end{aligned}
	\end{equation}
	where $\lambda(\mu)$ is defined in \eqref{eq:lambda}.
\end{lemma}
\begin{remark}
    We point out that in the second entry of \eqref{eq:Pi_At_2}, player $\mca$ secures the entire prize $\phi$ when endowed with the high budget $t=h$, even though player $\mcb$ invests non-zero resources to the competition (second entry of \eqref{eq:Bstar_2}). This is due to the players' equilibrium allocation profile $(\vec{F}_\mca^*, F_\mcb^*)$: in this regime, the support of $F_{\mca,j}^{h*}$ on the allocation to any battlefield $j$ is the interval $[2B,2(A_h-B)]\cdot \frac{v_j}{\phi}$, whereas the support of $F_{\mcb,j}^*$ is the interval $[0,2B]\cdot \frac{v_j}{\phi}$ (cf. Section A.5, the ``$\mcal{R}_4$" region \citep{Paarporn_2021_budget}). Thus, player $\mcb$ only competes with the low budget type $t=\ell$ in this case.
\end{remark}

In regimes where $\mcb$'s investment cost is low, player $\mca$'s payoff is decreasing in the belief $\mu$ (first entry of \eqref{eq:Pi_At_1},\eqref{eq:Pi_At_2}) regardless of the budget type. 

We now have characterizations to evaluate the objective function $\Pi_\mca(\sigma)$ for any $\sigma \in \Sigma_h$ \eqref{eq:final_payoff_Sigmah}. Before proceeding with the analysis, we state the following technical Lemma:
\begin{lemma}\label{lem:lambda_properties}
    The following properties hold.
    \begin{itemize}
        \item $\min\left\{\frac{\phi}{2\bar{A}(p)}, \lambda(p) \right\} = 
        \begin{cases}
            \lambda(p), &\text{if } p < \frac{A_h - 2A_\ell}{A_h - A_\ell} \\
            \frac{\phi}{2\bar{A}(p)}, &\text{if } p \geq \frac{A_h - 2A_\ell}{A_h - A_\ell}
        \end{cases}$
        \item For $p < \frac{A_h - 2A_\ell}{A_h - A_\ell}$, $\max\left\{\frac{\phi}{2\bar{A}(p)}, \lambda(p), \frac{\phi(1-p)}{2A_\ell} \right\} = \frac{\phi(1-p)}{2A_\ell}$.
        \item It holds that $\lambda(p)$ is strictly increasing on $p \in [0,\frac{A_h-2A_\ell}{A_h-A_\ell})$ and strictly decreasing on $p \in (\frac{A_h-2A_\ell}{A_h-A_\ell},1]$.
    \end{itemize}
\end{lemma}

\subsection{Proof of Theorem \ref{thm:signal_regions}}

\begin{figure*}[t]
    \centering
    \includegraphics[scale=0.16]{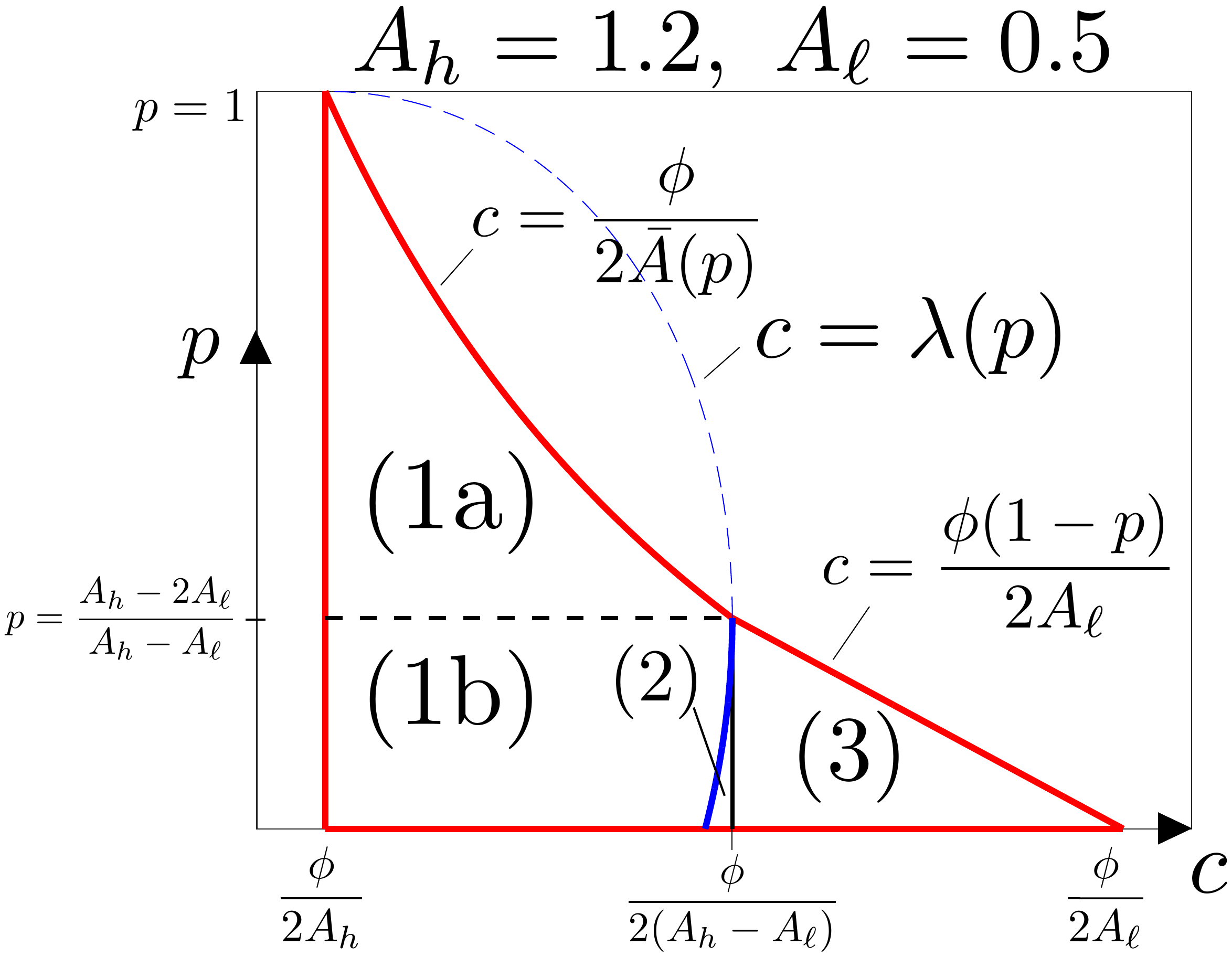}\hspace{2mm}
    \includegraphics[scale=0.16]{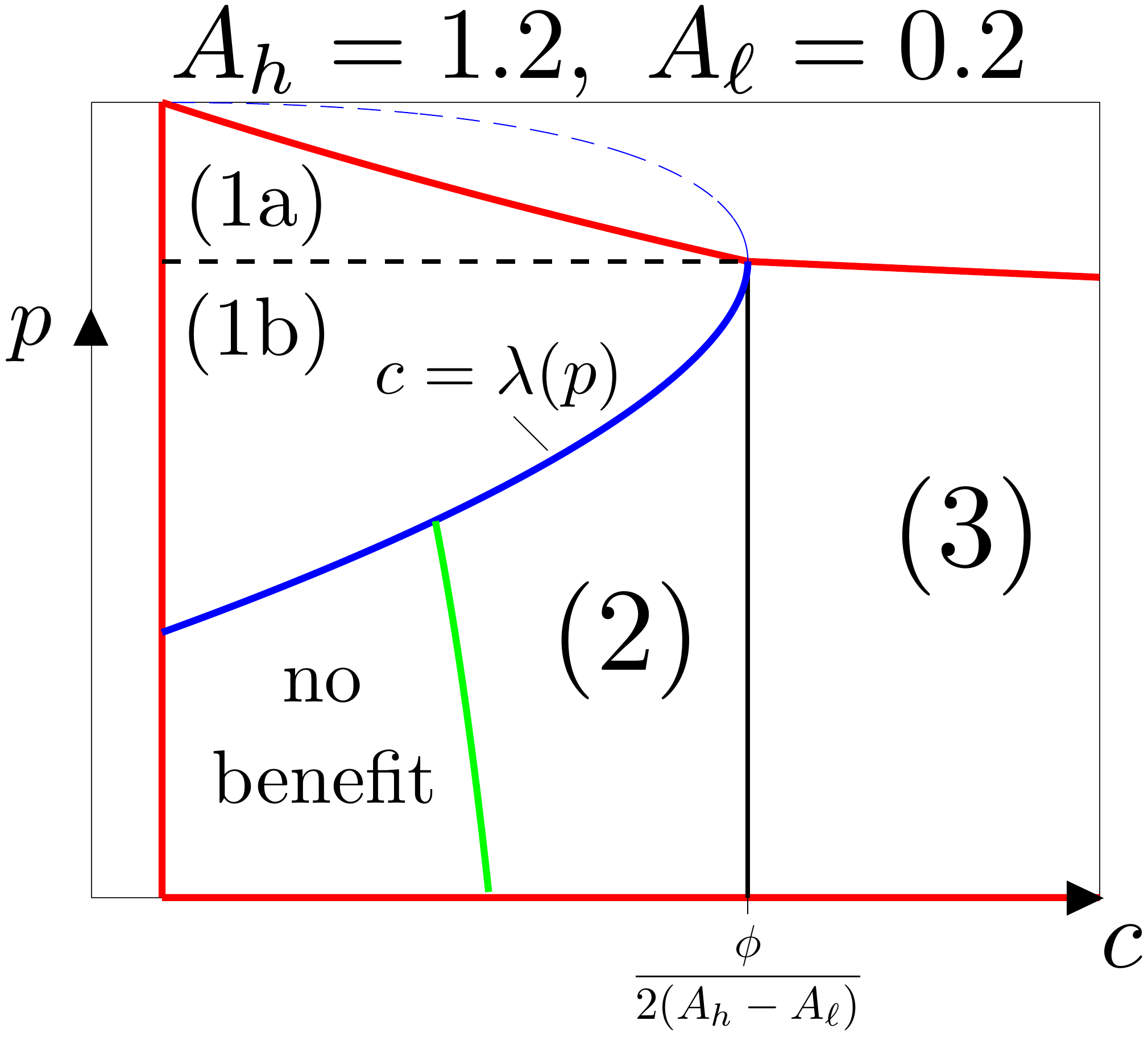}\hspace{2mm}
    \includegraphics[scale=0.16]{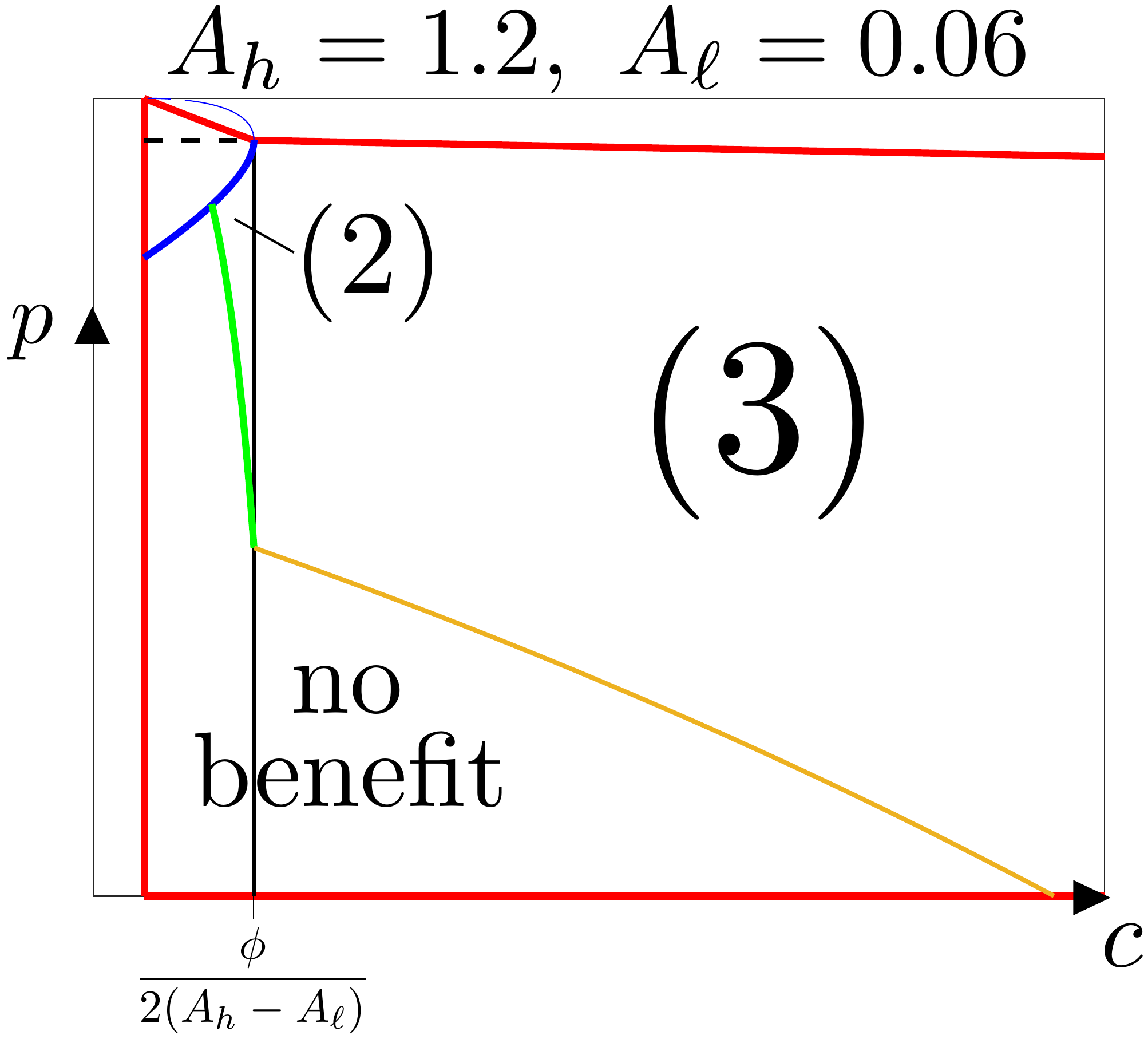}
    \caption{\small Diagrams that illustrate the regions in parameter space $(c,p)$ that correspond to each of the Cases in the proof of Theorem \ref{thm:signal_regions}. The regions of Cases 2 and 3 depend on the separation between high and low budgets $A_h,A_\ell$. (Left) There are benefits in signaling for the regions labeled 1a, 1b, 2, and 3. (Center) When $A_h = 1.2$ and $A_\ell = 0.2$, the condition \eqref{eq:case2_condition} is active in a portion of the Case 2 region (border indicated by the green line). (Right) When $A_h = 1.2$ and $A_\ell = 0.06$, a portion of the Case 3 region yields no benefit in signaling.}
    \label{fig:thm_pf}
\end{figure*}

We will split the proof into cases that correspond to the items in the statement of Theorem \ref{thm:signal_regions}. Figure \ref{fig:thm_pf} illustrates the parameter regions specified by these cases.

\noindent\underline{\textbf{Case 1a: } $\frac{\phi}{2A_h} \leq c < \frac{\phi}{2\bar{A}(p)}$ and $p \geq \max\left\{ \frac{A_h - 2A_\ell}{A_h - A_\ell},0 \right\}$}.

The payoff from a trivial policy is given by the first entry of \eqref{eq:Pi_ns_1}, $\Pi_\mca^\text{ns} = \sqrt{\frac{c\phi\bar{A}(p)}{2}}$. The belief $\mu_h = \frac{p}{p+q(1-p)}$ is strictly decreasing in $q$, and satisfies $\mu_h \geq p \geq \frac{A_h-2A_\ell}{A_h-A_\ell}$ for all $q \in [0,1]$. Thus,  $\pi_\mca(\P_h,B^*(\mu_h);t)$ is given by \eqref{eq:Pi_At_1}. Specifically,
\begin{equation}
    \pi_\mca(\P_h,B^*(\mu_h);t) = 
    \begin{cases}
        \phi, &\text{if } q \in \left[0, q_1^* \right] \\
        \sqrt{\frac{c\phi A_t^2}{2\bar{A}(\mu_h)}}, &\text{if } q \in \left(q_1^*,1 \right]
    \end{cases}
\end{equation}
where $q_1^* \triangleq \frac{p}{1-p}\frac{2cA_h-\phi}{\phi-2cA_\ell}$. Because $\mu_\ell = 0$ and $c < \frac{\phi}{2\bar{A}(p)} < \frac{\phi}{2A_\ell}$, $\pi_\mca(\P_\ell,B^*(\mu_\ell);\ell) = \sqrt{\frac{c\phi A_\ell}{2}}$ (either the first or second entry of \eqref{eq:Pi_At_2}). We can thus write $\Pi_\mca(q)$ \eqref{eq:final_payoff_Sigmah} as
\begin{equation}\label{eq:final_payoff_case1}
    \begin{aligned}
        &\phi(p + (1-p)q) + (1-p)(1-q)\sqrt{\frac{c\phi A_\ell}{2}}, \\
        &\quad\quad\text{for } q \in [0,q_1^*]. \\
        &\sqrt{\frac{c\phi}{2\bar{A}(\mu_h)}}\cdot (p A_h + (1-p)q A_\ell)+ (1-p)(1-q)\sqrt{\frac{c\phi A_\ell}{2}} \\  &\quad\quad\text{for } q \in (q_1^*,1]. \\
    \end{aligned}
\end{equation}
We first observe this is linearly increasing on $q \in [0,q_1^*]$. We proceed first by showing that $\Pi_\mca(0) > \Pi_\mca^\text{ns}$. Then, we will show that $\Pi_\mca(q) < \Pi_\mca^\text{ns}$ for $q \in (q_1^*,1]$, thus establishing that $\Pi_\mca(q^*) > \Pi_\mca^\text{ns}$ and $q_1^*$ is the optimal policy.

The claim that $\Pi_\mca(0) > \Pi_\mca^\text{ns}$ is equivalent to
\begin{equation}\label{eq:G}
    G(c,p) := p\phi + (1-p)\sqrt{\frac{c\phi A_\ell}{2}} - \sqrt{\frac{c\phi\bar{A}(p)}{2}} > 0
\end{equation}
We assert $G(c,p)$ is strictly increasing in $p$:
\begin{equation}
    \frac{\partial G}{\partial p} = \phi - \sqrt{\frac{c\phi}{2}}\left(\sqrt{A_\ell} - \frac{A_h-A_\ell}{2\sqrt{p_h(A_h-A_\ell) + A_\ell}} \right)
\end{equation}
If the term in the parentheses above is negative, then $\frac{\partial G}{\partial p} > 0$. If it is not, we can write the expression as $\frac{\partial G}{\partial p} = \phi - \sqrt{\frac{c\phi \hat{A}}{2}}$ for some $0 \leq \hat{A} < A_\ell$. Since $c < \frac{\phi}{2A_\ell}$, we have $\frac{\partial G}{\partial p} > 0$.

Let $p_\text{LB} := \max\{\frac{A_h-2A_\ell}{A_h-A_\ell},0\}$. If $p_\text{LB} = 0$, then from \eqref{eq:G}, $G(c,p_\text{LB}) = 0$. By the monotonicity of $G$, $G(c,p) > 0$ for any $p > 0$.

Now suppose $p_\text{LB} = \frac{A_h-2A_\ell}{A_h-A_\ell}$, implying $A_h \geq 2A_\ell$. We will show that $G(c,p_\text{LB}) > 0$ for all $c \in [\frac{\phi}{2A_h},\frac{\phi}{2\bar{A}(p_\text{LB})})$, where note that $\bar{A}(p_\text{LB}) \geq \bar{A}(p)$ for all $p \geq \frac{A_h - 2A_\ell}{A_h-A_\ell}$. By the monotonicity of $G$ in $p$, it will follow that $\Pi_\mca(0) > \Pi_\mca^\text{ns}(\P)$. We have
\begin{equation}
    \begin{aligned}
        G(c,p_\text{LB}) &= (A_h-2A_\ell)\phi + A_\ell\sqrt{\frac{c\phi A_\ell}{2}} \\ &\quad\quad\quad - (A_h-A_\ell)\sqrt{\frac{c\phi(A_h-A_\ell)}{2}}.
    \end{aligned}
\end{equation}

As a function of $c$, $G(c,p_\text{LB})$ is strictly decreasing:
\begin{equation}
    \frac{\partial G}{\partial c}(c,p_\text{LB}) = \frac{\sqrt{\phi}}{2\sqrt{2c}}\left(A_\ell^{3/2} - (A_h-A_\ell)^{3/2} \right) < 0
\end{equation}
It thus suffices to show that $G(\frac{\phi}{2(A_h-A_\ell)},p_\text{LB}) > 0$, or equivalently,
\begin{equation}
    \frac{1}{2}(A_h-A_\ell) - A_\ell\left(1 - \sqrt{\frac{A_\ell}{A_h-A_\ell}}\right) > 0
\end{equation}
whenever $A_h > 2A_\ell$. The above condition must hold: its value is positive when evaluated at $A_h = 2A_\ell$, and its derivative with respect to $A_h$ is $\frac{1}{2}(1 - (\frac{A_\ell}{A_h-A_\ell})^{3/2}) > 0$. Therefore, $G(c,p_\text{LB})>0$ for all $c \in [\frac{\phi}{2A_h},\frac{\phi}{2\bar{A}(p_\text{LB})})$, and it follows that $\Pi_\mca(0) > \Pi_\mca^\text{ns}$ for all $p \geq p_\text{LB}$.

We now proceed to show $\Pi_\mca(q) < \Pi_\mca^\text{ns}$ for $q \in (q_1^*,1]$. This claim follows by observing that $\Pi_\mca(1) =\Pi_\mca^\text{ns}$ and $\Pi_\mca(q)$ is strictly increasing in $q$: 
\begin{equation}
    \begin{aligned}
        \frac{\partial \Pi_\mca}{\partial q}(q) &\propto  \frac{p A_h+(1-p)qA_\ell + (p+(1-p)q)A_\ell}{2\sqrt{(p+(1-p)q)(p A_h+(1-p)A_\ell)}} - \sqrt{A_\ell} \\
        &= \frac{1}{2}\left(\sqrt{\bar{A}(\mu_h)} + \sqrt{\frac{A_\ell^2}{\bar{A}(\mu_h)}} \right) - \sqrt{A_\ell}
    \end{aligned}
\end{equation}
The above quantity is positive, since
\begin{equation}
    \begin{aligned}
        &\frac{1}{2}\left(\sqrt{\bar{A}(\mu_h)} + \sqrt{\frac{A_\ell^2}{\bar{A}(\mu_h)}} \right) > \sqrt{A_\ell} \\
        &\iff \left(\frac{A_\ell}{\sqrt{\bar{A}(\mu_h)}} - \sqrt{\bar{A}(\mu_h)} \right)^2 > 0.
    \end{aligned}
\end{equation}

\noindent\underline{\textbf{Case 1b: } $\frac{\phi}{2A_h} \leq c < \lambda(p)$ and $p < \max\left\{ \frac{A_h - 2A_\ell}{A_h - A_\ell},0 \right\}$}.

The proof in this regime is almost identical to the proof of Case 1a. The only difference is that $\mu_h < \frac{A_h-2A_\ell}{A_h-A_\ell}$ for some values of $q$. However, the resulting expression for $\Pi_\mca(q)$ is identical to \eqref{eq:final_payoff_case1}, and thus we omit these details.

\noindent\underline{\textbf{Case 2: } $\max\{\lambda(p),\frac{\phi}{2A_h}\} \leq c < \frac{\phi}{2(A_h-A_\ell)}$, $p < \frac{A_h - 2A_\ell}{A_h - A_\ell}$}.

The payoff from a trivial policy is given by the second entry of \eqref{eq:Pi_ns_2}, $\Pi_\mca^\text{ns} = p\phi + \sqrt{\frac{c\phi(1-p)A_\ell}{2}}$. The payoff $\pi_\mca(\P_\ell,B^*(\mu_\ell);\ell) = \sqrt{\frac{c\phi A_\ell}{2}}$ for all $q \in [0,1]$, since $\mu_\ell = 0$ and $c < \frac{\phi}{2A_\ell}$.

Thus, $\Pi_\mca(q)$ is given as follows. For $q \in [0,q_1^*]$, where $q_1^*$ was defined in Case 1a, it is
\begin{equation}
    \phi(p + (1-p)q) + (1-p)(1-q)\sqrt{\frac{c\phi A_\ell}{2}}.
\end{equation}
For $q \in (q_1^*,q_c]$, where $q_c$ satisfies $\lambda(q_c) = c$, it is
\begin{equation}
    \sqrt{\frac{c\phi}{2\bar{A}(\mu_h)}}(pA_h + (1-p)qA_\ell) + (1-p)(1-q)\sqrt{\frac{c\phi A_\ell}{2}}.
\end{equation}
For $q \in (q_c,1]$, it is
\begin{equation}
    p\phi + (1-p)q\sqrt{\frac{c\phi A_\ell}{2(1-\mu_h)}} + (1-p)(1-q)\sqrt{\frac{c\phi A_\ell}{2}}.
\end{equation}
Through a series of algebraic steps, we can verify that $\Pi_\mca(q) \leq \Pi_\mca^\text{ns}$ for $q \in (q_1^*,1]$ with equality if and only if $q = 1$. Now, observe $\Pi_\mca(q)$ is linearly increasing on $q \in [0,q_1^*]$. We then seek to identify parameters $c,p$ for which $\Pi_\mca(q_1^*) > \Pi_\mca^\text{ns}$. This condition is equivalent to \eqref{eq:case2_condition} from the Theorem statement.



\noindent\underline{\textbf{Case 3: } $\frac{\phi}{2(A_h-A_\ell)} \leq c < \frac{\phi(1-p)}{2A_\ell}$ and $p < \max\left\{ \frac{A_h - 2A_\ell}{A_h - A_\ell},0 \right\}$}.

Like in Case 2, we have $\Pi_\mca^\text{ns} = p\phi + \sqrt{\frac{c\phi(1-p)A_\ell}{2}}$, and $\pi_\mca(\P_\ell,B^*(\mu_\ell);\ell) = \sqrt{\frac{c\phi A_\ell}{2}}$ for all $q \in [0,1]$, since $\mu_\ell = 0$ and $\frac{\phi}{2(A_h-A_\ell)} < c < \frac{\phi}{2A_\ell}$.

Note that
\begin{equation}
    \frac{\phi}{2(A_h-A_\ell)} = \lambda(\frac{A_h-2A_\ell}{A_h-A_\ell}) > \lambda(p)
\end{equation}
for $p < \frac{A_h-2A_\ell}{A_h-A_\ell}$ (Lemma \ref{lem:lambda_properties}). 

Thus for $q \in [0,\hat{q}]$, $\pi_\mca(\P_h,B^*(\mu_h);t)$ is determined from \eqref{eq:Pi_At_1}. Observe that $c < \frac{\phi}{2\bar{A}(\mu_h)}$ is equivalent to $q > q_r \triangleq \frac{p}{1-p}\frac{2cA_h-\phi}{\phi-2cA_\ell}$. From the sub-case, it holds that $\hat{q} < q_r$, and therefore $\pi_\mca(\P_h,B^*(\mu_h);t) = \phi$ for $q \in [0,\hat{q}]$.

For $q \in (\hat{q},1]$, $\pi_\mca(\P_h,B^*(\mu_h);t)$ is determined from \eqref{eq:Pi_At_2}. From the sub-case condition and Lemma \ref{lem:lambda_properties}, $c > \lambda(\mu_h)$ for all $q \in (\hat{q},1]$. Moreover, $c > \frac{\phi(1-\mu_h)}{2A_\ell}$ is equivalent to $q < q_3^* \triangleq \frac{p}{1-p}\frac{2cA_\ell}{\phi-2cA_\ell}$. Then, $\pi_\mca(\P_h,B^*(\mu_h);t)$ is given by the third entry of \eqref{eq:Pi_At_2} for $q \in (\hat{q},q_3^*]$, and the second entry of \eqref{eq:Pi_At_2} for $q \in (q_3^*,1]$.

Thus, $\Pi_\mca(q)$ is:
\begin{equation}\label{eq:final_payoff_case3}
    \begin{aligned}
        &\phi(p + (1-p)q) + (1-p)(1-q)\sqrt{\frac{c\phi A_\ell}{2}} \\  
        &\quad\quad\text{for } q \in [0,q_3^*] \\
        &p\phi + (1-p)q\sqrt{\frac{c\phi A_\ell}{2(1-\mu_h)}} + (1-p)(1-q)\sqrt{\frac{c\phi A_\ell}{2}} \\ 
        &\quad\quad \text{for } q \in (q_3^*,1] \\
    \end{aligned}
\end{equation}
For the interval $q \in (q_3^*,1]$, it holds that $\Pi_\mca(q) \leq \Pi_\mca^\text{ns}$ with equality if and only if $q = 1$ (the same expression appears in Case 2). Now, observe that $\Pi_\mca(q)$ is linearly increasing on $q \in [0,q_3^*]$. We then seek to identify parameters $c,p$ for which $\Pi_\mca(q_3^*) > \Pi_\mca^\text{ns}$. This condition is equivalent to
\begin{equation}\label{eq:condition_3a}
    f_p > \frac{1}{2}\sqrt{\frac{\phi}{2cA_\ell}} \cdot \frac{\phi-2cA_\ell}{\phi - \sqrt{c\phi A_\ell/2} }
\end{equation}
where $f_p \triangleq \frac{p}{\sqrt{1-p} - (1-p)}$. The inequality \eqref{eq:condition_3a} can be re-written as 


\begin{equation}
    \sqrt{2\phi A_\ell}(f_p - 1) c - 2f_p\phi\sqrt{c} + \phi\sqrt{\frac{\phi}{2A_\ell}} > 0.
\end{equation}
The left-hand side above is quadratic in $\sqrt{c}$, with roots $r_\pm = \frac{\phi}{(f_p-1)\sqrt{2\phi A_\ell}}\left[f_p \pm \sqrt{f_p^2 - f_p + 1} \right]$. Thus, \eqref{eq:condition_3a} is equivalent to $c < r_-^2$ or $c > r_+^2$. However, we observe that $r_+^2 > \frac{\phi}{2A_\ell}$, so $c > r_+^2$ is a region outside the Case 3 region.

The cases we have analyzed above give sufficiency for the items in Theorem statement (and necessity for Cases 2 and 3). For all other parameters $(c,p)$ not considered in the cases, we observe:
\begin{itemize}[leftmargin=*]
    \item If $c < \frac{\phi}{2A_h}$, $\Pi_\mca^\text{ns} = \sqrt{\frac{c\phi\bar{A}(p)}{2}}$. $\Pi_\mca(q)$ is given by the expression in the second entry of \eqref{eq:final_payoff_case1} for all $q \in [0,1]$. We have already shown that $\Pi_\mca(q) \leq \Pi_\mca^\text{ns}$ with equality if and only if $q = 1$.
    \item If $c > \frac{\phi}{2\bar{A}(p)}$ and $p \geq \frac{A_h - 2A_\ell}{A_h - A_\ell}$, then $\Pi_\mca^\text{ns} = \phi$. Thus, no signaling is able to benefit $\mca$.
    \item If $c > \frac{(1-p)\phi}{2A_\ell}$ and $p < \frac{A_h - 2A_\ell}{A_h - A_\ell}$, then $\Pi_\mca^\text{ns} = \phi$. Thus, no signaling is able to benefit $\mca$.
\end{itemize}

\section{Conclusion}

This paper studies a competitive interaction between a signaler and an adversary, where the signaler has the opportunity to provide additional information about its capabilities to the adversary. We formulated this interaction as an extensive-form game, and used a General Lotto game model as the basis of the competition model. Leveraging recent results of incomplete information General Lotto games, we derived the optimal signaling policies within a sub-class of  policies. Moreover, we derived necessary and sufficient conditions under which the optimal policy offers performance improvements to the signaler over not signaling at all. Future work will focus on deriving optimal  policies over the entire space of signaling policies.


\bibliography{sources,library}             


\end{document}